\documentclass[
    ,final            
  ]
  {aipproc}

\layoutstyle{6x9}

\begin{document}
\title{Single Pion Measurement Capabilities at SciBooNE}


\author{Y. Nakajima for the SciBooNE collaboration}{
  address={Department of Physics, Kyoto University, Kyoto 606-8502, Japan}
}



\begin{abstract}
 The precise knowledge of the single pion production
 cross-section of neutrino around the $\sim$1 GeV energy region is an
 essential ingredient in the interpretation of neutrino oscillation experiments.
 The unique opportunities and prospects of single pion measurements at
 SciBooNE are described.
\end{abstract}

\maketitle


 \section{Overview}
 The SciBooNE experiment\cite{sciboone-prop} is a new
 neutrino experiment at FNAL
 aiming for precise cross-section
 measurements around the $\sim$1 GeV energy region.

 The detector is located on the Booster Neutrino Beamline (BNB)
 at Fermilab, which has been operated since 2002 for the MiniBooNE
 experiment\cite{AguilarArevalo:2007it}.
The property of this beamline is well understood by MiniBooNE and also
thanks to precise hadron production measurements by HARP\cite{harp-2007}.
The mean neutrino energy is $\sim$0.8 GeV, which is almost the same as that
of T2K\cite{t2k-loi}.
  The detailed description of SciBooNE detector complex can be found
  elsewhere\cite{sciboone-prop,rob}.
  The main component, SciBar\cite{nitta-2004},
  is a fully active finely segmented tracking
  detector which previously used at K2K.
  This combination of SciBar and the well understood high intensity BNB 
  will give us great
  opportunities for precise cross-sections measurements around the $\sim$1
  GeV.
  The cross-section data in this energy range will
  directly help T2K understand their neutrino flux and
  also MiniBooNE as a near detector.
  
  \section{Motivations to study single pion productions}

  Understanding the cross-section of charged current single pion
  production (CC1$\pi^+ = \nu_{\mu} N \to \mu^- N \pi^+$, where N=p,n) is especially
  important to study $\nu_{\mu} \to \nu_X$ oscillation, which will
  be made by, for example, T2K and MiniBooNE.
  These experiments utilize Cherenkov type detectors for neutrino
  detection and the signal event will be generated by muon
  from charged current quasi
  elastic scattering (CCQE) because it is possible to
  reconstruct the energy of neutrino from observed muon.
  However, this measurement suffers from large number of backgrounds
  which dominantly come from 
  CC1$\pi^+$ interactions.
  This is because the pion is not observed since its momentum is below the
  Cherenkov threshold. Moreover, the pion can be absorbed in nuclei,
  making
  the distinction between signals and backgrounds even harder to accomplish.

  In the case of T2K, the CC1$\pi^+$/CCQE cross-section ratio is required to
  be understood at 5\% level 
  to keep the resulting error on the oscillation parameters
   comparable to that due to statistical uncertainties\cite{sciboone-prop}.
  However, the past cross-section measurements have been limited by low
  statistics and large systematic uncertainty. Therefore, precise measurements
  are strongly needed.
  Recently, K2K and MiniBooNE are
  starting to make precision measurements on 
  CC1$\pi^+$ cross-section\cite{bonnie,lisa},
  and SciBooNE will follow with higher
  statistics than K2K and better resolution than MiniBooNE.
  
  \section{CC1$\pi^+$ measurement at SciBooNE}
  \subsection{Detector Performance}
\begin{figure}[!ht]
   \includegraphics[height=2.5cm,clip,trim=0 0 0 0]{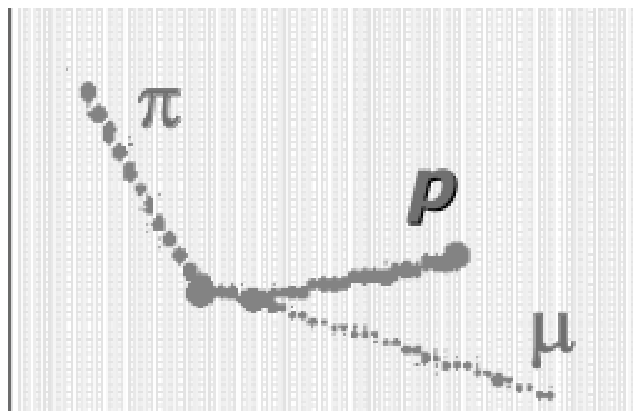}
   \includegraphics[height=2.5cm,clip,trim=0 0 10 0]{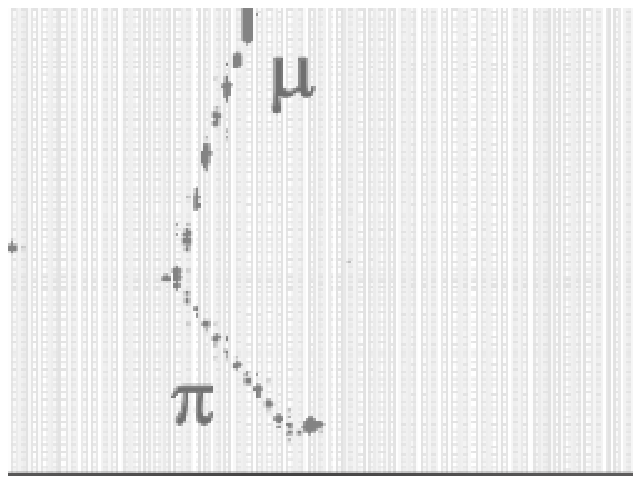}
   \includegraphics[height=2.5cm,clip,trim=30 0 0 0]{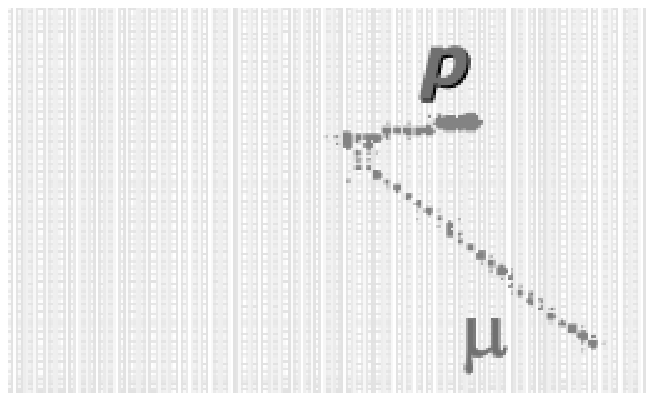}
   \caption{Event displays of typical Monte Calro CC1$\pi^+$ interactions.}
   \label{fig:mcdisp}
  \end{figure}

  One of the most important features of the SciBar detector
  is that every charged particle
  including the protons and pions can be reconstructed as tracks.
  Figure~\ref{fig:mcdisp} shows Monte Carlo event displays for CC1$\pi^+$
  interactions occurring in the SciBar detector.

  The event display on the left is a typical $\nu_{\mu} p \to \mu^- p \pi^+$ event.
  We can see the tracks of all charged particles and also can clearly
  distinguish the muon/pion tracks from the proton track
  by their energy depositions. So we can clearly identify such an event with two
  MIP-like tracks and a proton-like track.

  Also, we can distinguish $\nu_{\mu} n \to \mu^- n \pi^+$ from
  $\nu_{\mu} p \to \mu^- p \pi^+$ by the existence of a proton track.
  The center figure shows a typical $\nu_{\mu} n \to \mu^- n \pi^+$
  event.
  Furthermore, even if the pion is absorbed in the nucleus, we can distinguish
  CC1$\pi^+$ from CCQE by using proton and muon kinematics. The right
  figure is an example of $\nu_{\mu} p \to \mu^- p \pi^+$ where the
  $\pi^+$ is absorbed.

  This clear event-by-event final state tagging should enables us to
  unambiguously identify CC1$\pi^+$ event, and also to study kinematical
  characteristic such as pion momentum distribution or pion
  absorption cross-section.

  \subsection{Current Status and Expected Sensitivity}
\begin{figure}[!ht]
  \includegraphics[angle=90,width=8cm]{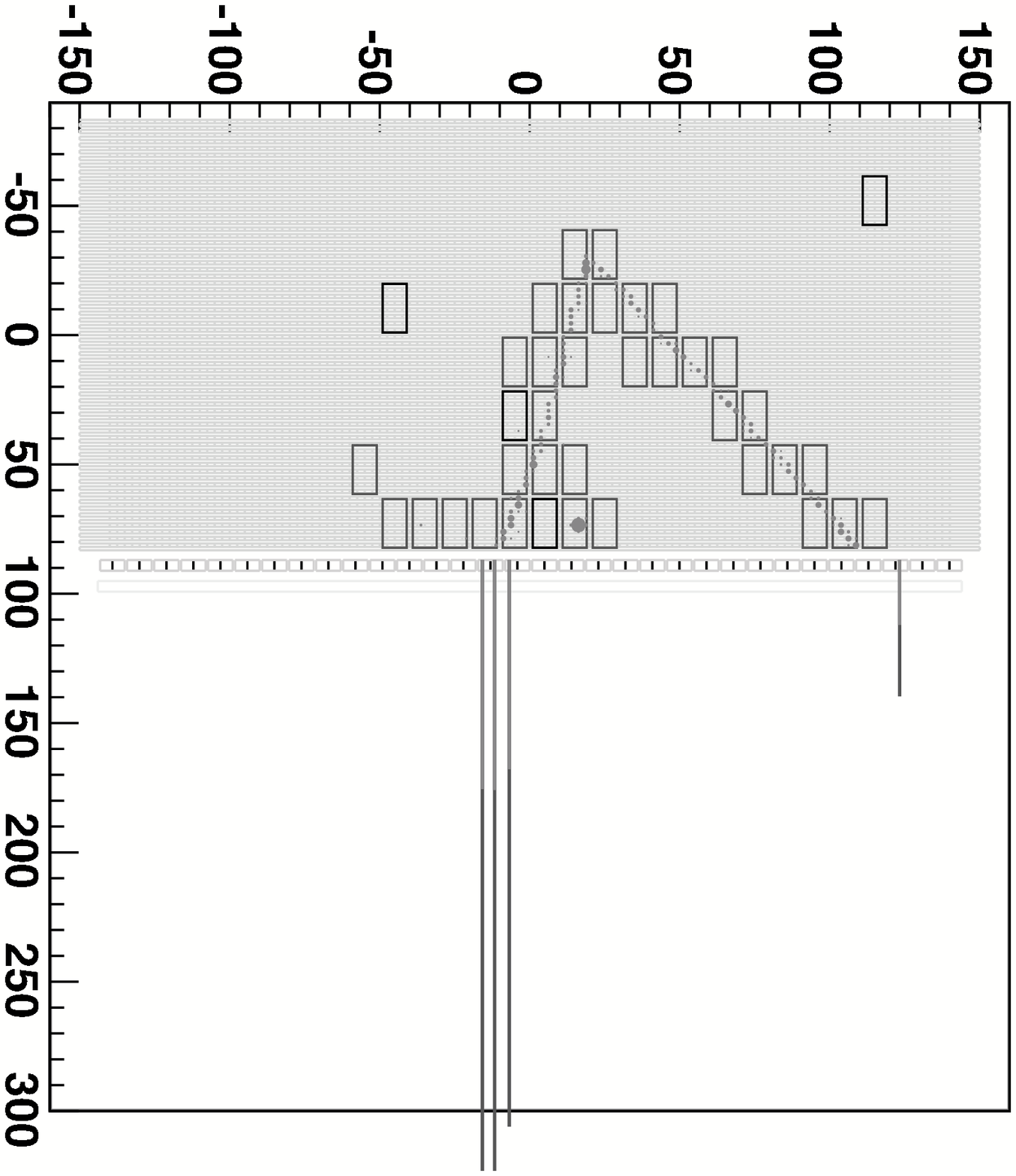}
  \caption{An event display of a
  $\stackrel{\scriptscriptstyle (-)}{\nu}_{\mu} p \to \mu^{\pm} p \pi^{\mp}$
  candidate.}
  \label{fig:disp}
 \end{figure}

  We completed detector construction and installation in April 2007.
  Then after the detector commissioning using cosmic rays, we started
  the commissioning using the BNB at the
  end of May 2007, and we successfully detected (anti-)neutrino events.
  Figure~\ref{fig:disp} is an event display of a
  $\stackrel{\scriptscriptstyle (-)}{\nu}_{\mu} p \to \mu^{\pm} p \pi^{\mp}$
  candidate, which we took in the early
  stage of the commissioning. One can clearly see two MIP-like tracks
  which come from a muon and a pion, and the vertex activity which
  results from a proton.
 
 The commissioning phase allowed us to confirm that the detectors are working as
  expected and we are collecting (anti-)neutrino data.
  It is expected to collect $\sim$28000 CC1$\pi^+$ interactions
  with $1\times 10^{20}$protons on target for neutrino mode.
  This allows us to measure
  CC1$\pi^+$/CCQE cross-section ratio with less than 5 \% statistical
  uncertainty, which will provide sufficient input for
  $\nu_{\mu}$ disappearance measurements at T2K and MiniBooNE.


\begin{theacknowledgments}
 SciBooNE collaboration gratefully acknowledges support from various
 grants and contracts from the Department of Energy (U.S.), the National
 Science Foundation (U.S.), the MEXT (Japan), the INFN (Italy) and the
 Spanish Ministry of Education and Science.
 The author was supported by Japan Society for the Promotion of Science.
\end{theacknowledgments}

\end{document}